\documentstyle[12pt]{article}

\def\beq{\begin{equation}}
\def\eeq{\end{equation}}

\def\bea{\begin{eqnarray}}
\def\eea{\end{eqnarray}}
\def\ben{\begin{enumerate}}
\def\een{\end{enumerate}}

\def\ni{\noindent}
\def\nn{\nonumber}

\def\tr{\mbox{Tr}\, }
\def\brr{\begin{array}}
\def\err{\end{array}}

\def\e{{\rm e}}
\def\nn{\nonumber \\}

\begin{document}

\begin{flushright}
CEAB 95/9-13 \\
NDA-FP-22 \\
OCHA-PP-66 \\
hep-th/9511167 \\
August 1995 \\
\end{flushright}

\vspace{2mm}


\begin{center}

{\LARGE \bf
Quantum effects \\
of stringy and membranic nature \\
for the swimming of micro-organisms \\
in a fluid}

\vspace{4mm}

{\sc E. Elizalde$^\diamondsuit$}\footnote{
E-mail: eli@zeta.ecm.ub.es},
{\sc M. Kawamura$^\heartsuit$}\footnote{
E-mail: masako@phys.ocha.ac.jp},
{\sc S. Nojiri$^\spadesuit$}\footnote{
E-mail: nojiri@cc.nda.ac.jp}, \\
{\sc S.D. Odintsov$^\clubsuit$}\footnote{
E-mail: sergei@ecm.ub.es.} and
{\sc A. Sugamoto$^\heartsuit$}\footnote{
E-mail: sugamoto@phys.ocha.ac.jp}\\

\

$\diamondsuit$ Dipartimento di Fisica,
Universit\`a degli Studi di Trento, Italia \\
and Center for Advanced Studies CEAB, CSIC, Cam\'{\i}
de Santa B\`arbara,
17300 Blanes, Spain

\

$\heartsuit$ Department of Physics, Ochanomizu University\\
1-1, Otsuka 2, Bunkyo-ku, Tokyo, 112, Japan \\

\

$\spadesuit$ Department of Mathematics and Physics,
National Defence Academy\\
Yokosuka, 239, Japan

\

$\clubsuit$ Tomsk Pedagogical Institute,
634041 Tomsk, Russia, \\
and Department ECM, Faculty of Physics,
University of  Barcelona, \\
Diagonal 647, 08028
Barcelona, Spain

\end{center}

\

\begin{center}
{\bf Abstract}
\end{center}

The static potential is investigated in  string
 and membrane theories
coupled to $U(1)$ gauge fields (specifically, external
magnetic fields) and with antisymmetric tensor fields.
The explicit dependence of the potential on the shape of the
extended objects is obtained, including a careful calculation of
 the quantum  effects.
Noting the features which are common to the dynamics of strings
and membranes  moving in background fields and to
the swimming of micro-organisms  in a
fluid, the latter problem is studied. The
Casimir energy of a
micro-organism is estimated, taking into account the quantum effects
and the  backreaction from the outside fluid.

\vfill

\newpage

 \ni{\bf 1. \ Introduction.}


The very basic problem in Biology of the description of
the swimming of
micro-organisms has been analysed from a gauge
theoretical point of view by Shapere and
Wilczek \cite{c6,c7} and from a string (or membrane)
theoretical viewpoint in Refs. \cite{c8}-\cite{c12}.
The surface of a micro-organism has one (resp. two) dimensions
in a two (resp. three) dimensional fluid, and it can be
regarded as
a special kind of string (or membrane). In particular, the shapes
of the surfaces of  ciliate and  flagellate in two
dimensions are similar to those of closed and open
strings, respectively.
Owing to the backreaction of the velocity fields, there
appear interactions between these strings,
which are similar to the interaction mediated by
an anti-symmetric tensor field \cite{eo1} and others.

In the case of micro-organisms, the string (i.e.,
the surface of the micro-organism) couples to the
velocity fields in a special way. Furthermore the
velocity fields can be regarded as gauge
fields in a $U(1)$ gauge theory. By integrating the velocity
fields,  Coulomb-like interactions are created between the strings.
Therefore, it is an interesting issue to consider in
a unified fashion the two
 problems, namely that of  string and membrane theories coupled
with  various  external fields, including the $U(1)$
gauge field, and  the swimming
problem of micro-organisms inside a fluid,
and thus obtain the static potential and the Casimir energy
corresponding to the last one from our knowledge of the
 first, general theory.

The paper is organized as follows. In sections 2 and 3
the theoretical background for the analysis will be
provided.
In particular, in Sect. 2 we obtain the static potential
for a bosonic string in a magnetic field.
In Sect. 3 the formulas for the case of the bosonic membrane
in a magnetic field are derived.
In section 4, the estimation of the static potential in
the string theory coupled to an antisymmetric field
is briefly summarized.
In Sect. 5, we proceed with the application of the theory
to the description of the swimming of micro-organisms.
In particular, we consider in detail
the quantum (thermal)
effects ---after adding the potential term--- and we
show explicitly the appearance of a kind of Casimir energy.

Of course, there are some differences between the usual string
and the micro-organism considered as such. The energy of the usual
string is proportional to its length  but
the stable shape of ciliate should be a circle with a
finite  radius and the
flagellum is not elastic (the length is not changable).
Another difference is that the apparent divergence in the
zero-point energy is not an ultra-violet but an infrared one.
The reason for this is the fact that
the efficiency of the translation or  rotation
of the micro-organism becomes better for higher
modes \cite{c7}.

\vspace{3mm}

\ni{\bf 2. \ Static potential for the bosonic string in
a magnetic field.}

We will consider here a bosonic string theory interacting
with a magnetic field \cite{c1}. The corresponding action
has the following form (in the Euclidean notation)
\beq
S= \frac{k}{2} \int d^2\xi \, \sqrt{g} \, \left[
\partial_i X^\mu  \partial^i X_\mu
+2 A_p (X)
\partial_i X^p  \partial^i X^1\right],
\label{n1}
\eeq
being $i=0,1$ and where the string coordinates $X^\mu$
describe a four-dimensional space-time ($ \mu =0,1,p$,
with $p=2,3$).
The parametrization of the string coordinates is
\beq
X^0_{cl} = \xi_0, \qquad
X^1_{cl} = \xi_1, \qquad
0 \leq \xi_0 \leq t, \
0 \leq \xi_1 \leq R,
\eeq
with the corresponding quantum fluctuations vanishing at
the boundary.
This represents a once-wrapped torus.

The semi-classical quantization of the theory is
conveniently considered in the background gauge
\beq
X^0 = X^0_{cl}, \qquad X^1 = X^1_{cl},
\eeq
in which there are no Faddeev-Popov ghosts. At the
one-loop level, the only quantum fluctuations which
contribute to the path integral are
those corresponding to $X^3$ and $X^4$ \cite{c2,c3}.

The following form will be chosen for the vector field:
\beq
A_p(X)= - \frac{1}{2}f \epsilon_{pm}X^m.
\label{n4}
\eeq
Note that $f$ in Eq. (\ref{n4}) plays the role of a
magnetic field. In principle one can consider the more
general situation when $D >4$, by
writing the vector field in a block-diagonal form,
being each block described as in (\ref{n4}).

The static potential or Casimir energy at one-loop can
be obtained as follows:
\beq
V = - \lim_{t \rightarrow \infty} \frac{1}{t} \ln \int {\cal D} X^m \,
e^{-S(t)}, \label{n5}
\eeq
where
\beq
S (t)= k \int_0^t d\xi_0 \int_0^R d\xi_1  \left[
1 + \frac{1}{2} \partial_i X^m \partial^i X_m
 - \frac{f}{2} \left(
X^3 \partial_1 X^2- X^2 \partial_1 X^3 \right) \right].
\label{n6}
\eeq
Performing the integrations over $X^2$ and $X^3$, by
taking into account the fact that for those fields
the classical background is zero, and by
satisfying the standard periodic boundary conditions,
one obtains
\beq
V=  kR + \frac{1}{2} \, \tr \ln \left( \brr{cc}
 - \Box & f\partial_1
\\ -f \partial_1 & - \Box \err \right)
=  kR + \frac{1}{2} \, \ln \det \left( \Box^2
+ f^2\partial_1^2
\right),
\label{n7}
\eeq
where $\Box = \partial_0^2 + \partial_1^2$. Observe that
when we write the operators Tr ln and ln det we understand
that the factor $1/t$ is included. In absence of
the magnetic field, $f=0$, the calculation of
 ln det $\Box$ yields \cite{c2,c3}
\beq
V (R)=  kR - \frac{\pi}{12R}.
\label{n8}
\eeq
Notice again that this is the one-loop result. In the case
of a non-vanishing magnetic field $f\neq 0$, using the zeta-function method
\cite{zfr1,dowk1} (for review,see \cite{c4}) we get \cite{c5}:
\beq
V (R)=  kR - \frac{1}{2} \, \zeta_A' (0),
\label{n9}
\eeq
where
\bea
\zeta_A (s) &=& \int d^2p \ (p\cdot p + f^2p_1^2)^{-s} = \int_0^\infty
dp \sum_{n=1}^\infty
\left[ p^2 + \frac{(1+f^2)n^2}{R^2} \right]^{-s}
\label{n10} \\  &
= & \frac{1}{\Gamma (s)}  \int_0^\infty dt \, t^{s-1}
\sum_{n=1}^\infty
e^{- (1+f^2)
n^2 t/R^2} \int_0^\infty
dp \ e^{-p^2 t} = -
\frac{\sqrt{\pi} \Gamma (s-1/2) R^{2s-1}}{2\,\Gamma (s)\,
(1+f^2)^{s-1/2}} \, \zeta (2s-1). \nn
\eea
By taking now the derivative at $s=0$, this yields the result
\beq
 -\zeta_A'(0) = \int d^2p \, \ln (p\cdot p +f^2p_1^2)
= \int_0^\infty dp
\sum_{n=1}^\infty
\ln \left( p^2 + \frac{(1+f^2)n^2}{R^2} \right) =
 - \frac{\pi \,
\sqrt{1+f^2}}{6R}. \label{n11}
\eeq
Finally, for the static potential we obtain
\beq
V (R)=  kR - \frac{\pi \, \sqrt{1+f^2}}{12R}.
\label{n12}
\eeq
The stronger the magnetic field is, the bigger is the quantum
correction to the static potential.

\vspace{3mm}

\ni{\bf 3. \ Bosonic membrane with a magnetic field.}

Let us now consider the action corresponding to the
Howe-Tucker bosonic membrane \cite{c11}. In the
Euclidean notation, it is given by the action
\beq
S= \frac{k}{2} \int d^3\xi \, \sqrt{\gamma} \,
\left( \gamma^{ij}
\partial_i X^\mu  \partial^j X_\mu -1
\right),
\label{nb1}
\eeq
where $k$ is the membrane tension. Here $\xi^i$, $i=0,1,2$,
denote the world-volume coordinates,
$\gamma^{ij}$  the world-volume metric, and $ X^\mu$,
$\mu = 0,1, \ldots, d-1$, denote the space-time
coordinates, with metric $g^{\mu\nu}$. In particular,
we will consider here the  situation
corresponding to $d=5$ space-time coordinates, that is
$\mu =0,1,2,p$, with $p=3,4$.

The classical solution for the theory (\ref{nb1}) of
toroidal form is given by
\beq
X^0_{cl} = \xi_0, \ \ \ X^1_{cl} = \xi_1, \ \ \ X^2_{cl}
= \xi_2,
\label{nb2}
\eeq
and $(\xi_1,\xi_2) \in R \equiv [0,A_1] \times [0,A_2]$.
It represents a once-wrapped torus.

For the semi-classical quantization of the theory, it
is convenient to consider it in the background gauge
\beq
X^0 =X^0_{cl} , \ \ \ X^1=X^1_{cl}, \ \ \ X^2=X^2_{cl}.
\label{nb3}
\eeq
In this gauge there are again no Faddeev-Popov ghosts, in
other words, quantum fluctuations for the three
space-time coordinates above should not be taken
into account in the calculation of the
corresponding path-integral. For $X^p$,
$p=3,4$, there are no classical fields.
They represent the fluctuations that
are completely quantum, and satisfy the standard
periodic boundary conditions
corresponding to the torus.

We can now consider the interaction of the bosonic
membrane with a vector field:
\beq
S_I= \frac{k}{2} \int d^3\xi \, \sqrt{\gamma} \, \left[2A_p(X)
\partial_i X^p \left( \partial^i X^1
+  \partial^i X^2 \right) \right].
\label{nb4}
\eeq
The form of the vector field can be conveniently chosen
to be the following
\beq
A_p (X)= -\frac{1}{2} f\epsilon_{pm} X^m,
\label{nb5}
\eeq
where $f$ plays the role of the magnetic field. It
is interesting to observe that the interaction of a
string theory with a magnetic field  can be
considered in a similar way (see the previous section).

The static potential can be obtained as
\beq
V= -\lim_{t \rightarrow \infty} \frac{1}{t} \ln \int
{\cal D} X^m \, e^{-S(t)},
\label{nb6}
\eeq
where
\beq
S(t)= k \int_0^t d\xi_0 \int_R d^2\xi \left[ 1
+ \frac{1}{2} \partial_i X^m
\partial^i X_m + \frac{f}{2}  X^3 \left( \partial_1 X^4
+ \partial_2 X^4
\right) - \frac{f}{2}  X^4 \left( \partial_1 X^3
+ \partial_2 X^3
\right) \right].
\label{nb7}
\eeq

Now, one can calculate the path integral over $X^m$.
In the one-loop approximation, we get
\beq
V= k A_1A_2 + \frac{1}{2} \, \tr \ln \left( \brr{cc}
 - \Box & f(\partial_1 +
\partial_2) \\ -f (\partial_1 + \partial_2) & - \Box
\err \right),
\label{nb8}
\eeq
that is,
\beq
V =  k A_1A_2 + \frac{1}{2} \, \ln \det \left[ \Box^2
+ f^2 (\partial_1 + \partial_2)^2
\right],
\label{nb9}
\eeq
where $\Box = \partial_0^2 + \partial_1^2 + \partial_2^2$
is here the three-dimensional d'Alembertian.

As before, in the general case of a
 non-vanishing magnetic field, $f\neq 0$, using the zeta-function method
 \cite{c4} we obtain \cite{c5}:
 \beq
 V (A_1,A_2)=  kA_1A_2 - \frac{1}{2} \, \zeta_A' (0),
 \label{1n9}
 \eeq
 where the zeta function is now given by
 \bea
 \zeta_A (s) &=& \int d^2p \ [p\cdot p + f^2(p_1+p_2)^2]^{-s}\nn \\ & =&
\int_0^\infty
 dp \sum_{n_1,n_2}
 \left[ p^2 + (1+f^2) \left( \frac{n_1^2}{A_1^2}+ \frac{n_2^2}{A_2^2} \right) +
2f^2
\frac{n_1n_2}{A_1A_2} \right]^{-s}.  \nn \\  &
  = &
 \frac{\sqrt{\pi} \Gamma (s-1/2)}{4\,\Gamma (s)} \, F \left(s-\frac{1}{2};
\frac{1+f^2}{A_1^2},
 \frac{2f^2}{A_1A_2}, \frac{1+f^2}{A_2^2} \right), \label{1n10}
 \eea
The function $F(s;a,b,c)$ is the analytical continuation of the general
Epstein zeta function
in two dimensions, which is given by the celebrated Chowla-Selberg
formula \cite{cs} (for extensions
of this formula to inhomogeneous cases and truncated sums, see
\cite{eecs1,eecs2})
\begin{eqnarray}
&& F(s;a,b,c) = 2\zeta (2s)\, a^{-s} + \frac{2^{2s}
\sqrt{\pi}\, a^{s-1}}{\Gamma (s) \Delta^{s-1/2}} \,\Gamma (s
 -1/2) \zeta (2s-1) + \frac{2^{s+5/2} \pi^s }{\Gamma (s)
\Delta^{s/2-1/4}\sqrt{a}}
\nn \\ && \hspace{1cm} \times \sum_{n=1}^\infty n^{s-1/2} \, \sigma_{1-2s} (n)
\,
 \cos (n \pi b/a)  \, K_{s - 1/2}\left( \frac{\pi n \sqrt{\Delta}}{a} \right).
\label{cs1}
\end{eqnarray}
where
\begin{equation}
\sigma_s(n) \equiv \sum_{d|n} d^s,
\end{equation}
this sum extending over all divisors $d$ of $n$. In the general theory
dealing
with the homogeneous case, one assumes that $a,c >0$ and that
the discriminant
\beq
\Delta =4ac-b^2 >0
\eeq
(see \cite{cs}).
We obtain
\beq
\Delta = \frac{4}{A_1^2A_2^2} (1+f^2+f^4)
\eeq
and for the zeta function
 \bea
 \zeta_A (s) &=& \frac{\sqrt{\pi}}{2} \frac{(1+f^2)^{1/2-s}}{A_1^{1-2s}} \frac{
 \zeta (2s-1)}{\Gamma (s)} + \frac{2^{2s-3}
\pi\, A_1^{3-2s}}{A_1^{3-2s} \Delta^{s-1}} \, \frac{ \zeta (2s-2)}{s-1}
\label{1cs1} \\ && \hspace{-8mm} + \frac{(2\pi)^s A_1 }{\Gamma (s)
\Delta^{(s-1)/2}\sqrt{1+f^2}}
\sum_{n=1}^\infty n^{s-1} \, \sigma_{2-2s} (n) \,
 \cos \left(2 \pi n \frac{A_1}{A_2} \frac{f^2}{1+f^2}\right)  \, K_{s -
1}\left( \frac{\pi n
A_1^2 \sqrt{\Delta}}{1+f^2} \right).
\nn
\end{eqnarray}
The good convergence properties of expression (\ref{cs1})
 were very much
prised by Chowla and Selberg. They render the use of the
formula (\ref{1cs1}) above quite
simple. In fact, the two first terms are just nice  while
the last one (impressive in appearence) is very quickly
convergent  in the real case,
and thus absolutely harmless in practice. Only a few
first terms of the three series of Bessel functions in (\ref{1cs1}),
 need to be
calculated, even if one demands good accuracy.

 By taking now the derivative at $s=0$, this zeta function yields the
following result for the effective potential
 \bea
  V (A_1,A_2) &=&  kA_1A_2  - \frac{\pi \,
 \sqrt{1+f^2}}{12A_1} + \frac{\pi \zeta'(-2)A_1^3\Delta}{8 (1+f^2)^{3/2}}
\label{1n11} \\
&& - \frac{ A_1 \sqrt{\Delta} }{\sqrt{1+f^2}}
\sum_{n=1}^\infty n^{-1} \, \sigma_{2} (n) \,
 \cos \left(2 \pi n \frac{A_1}{A_2} \frac{f^2}{1+f^2}\right)  \, K_{s -
1}\left( \frac{\pi n
A_1^2 \sqrt{\Delta}}{1+f^2} \right). \nn
 \eea
As one can see, the $f$-dependence of the static potential is much more
complicated, as compared with string theory.

\vspace{3mm}

\ni{\bf 4. \ Static potential for the string coupled to an
 antisymmetric tensor field.}

 This case has been studied in \cite{eo1}. After
integrating over $A_{ij}$ and $\phi$ in order to obtain an
effective theory for the closed bosonic string, exhibiting a
Coulomb-like interaction term (compare with Ref. \cite{4}), we get
\bea
S&=& \int d^2\xi \, \sqrt{g} \, \left[ \frac{1}{2} G_{\mu\nu} (X)
g^{ij} \partial_i X^{\mu}  \partial_j X^{\nu} \right] + S_{int},
\nn \\
S_{int}&=& \int d^2\xi d^2\xi' \, \left[ c_1 e^2 \sigma_{\mu\nu} (\xi
) \sigma^{\mu\nu} (\xi' ) + c_2 \zeta^2 {R^{(2)}}^2 \right] V(|x-
x'|),
\label{555}
\\ &&
\sigma_{\mu\nu} (\xi ) =  \epsilon^{ij}  \partial_i X^{\mu}
\partial_j X^{\nu}, \ \ \ \  V(|x-x'|) =\frac{1}{|x(\xi )- x(\xi'
)|^2+a^2}, \nn
\eea
being $c_1$ and $c_2$ non-essential numerical constants which can
be choosen to be equal to 1.
The static potential in string theory is an interesting magnitude
in connection with the possible applications of string theory to QCD.
This fact was realized long ago \cite{6}, and a calculation of
the static potential in different string models has been carried
out explicitly in Refs. \cite{c2,c3}, and in particular for the
rigid string \cite{8} in Refs. \cite{9}-\cite{11}. It has been
pointed out in those works that the leading corrections to
the static potential have a universal character \cite{c3}.
In the applications of the static potential to QCD one can usually
choose the Wilson loop $C$ to be a rectangle on the plane, of
length $T$ and width $R$ (with $T>> R$). Then, the loop
expectation value can be found as follows \cite{12}:
\beq
W[C] \sim \exp [-TV (R)].
\label{556}
\eeq
The explicit one-loop calculation for a standard bosonic string
gives:
\beq
V(R) = kR + \frac{D-2}{2} \tr \ln \Box,
\label{557}
\eeq
where $k$ is again the string tension and $\Box$ the
two-dimensional d'Alembertian in the space of topology $R\times
S^1$. Using the zeta-function regularization procedure to
calculate (\ref{557}) one obtains the well-known result
\cite{c3,c2,c5} (for simplicity, we consider the case $D=4$)
\beq
V (R)= k R - \frac{(D-2)\pi}{24R}.
\label{558}
\eeq
The calculation to one-loop of the static
potential (\ref{557}) taking into account the Coulomb-like term
induced by the antisymmetric tensor fields, as in (\ref{555}),
 is quite non trivial since
 the integration over the $X^{\mu}$s is not
Gaussian.
The natural way to consider the integration is by decomposing the
variables as follows:
\beq
X^{\mu}(\xi) = X_0^{\mu}(\xi) + X_1^{\mu}(\xi),
\label{559}
\eeq
where $ X_0^{\mu}(\xi)$ is a background variable, which is a linear
function of $\xi$ satisfying the field equations
\beq
X_0^{\mu} (\xi) = c_i^\mu \xi^i, \qquad \eta_{\mu\nu} c_i^\mu c_j^\nu
=\eta_{ij},
\label{5510}
\eeq
being  the $c_i^\mu$ some  constants. The expansion of Eq.
(\ref{555}) up to second order on the fluctuations
$X_1^\mu(\xi)$ can be performed in the same way as it was
done in Ref. \cite{4}. After the subsequent functional Gaussian
integration over the $X_1^\mu$'s, we obtain the static potential
\beq
V(R)= kR + \frac{D-2}{2} \int dp
\sum_{n=1}^\infty \ln \left[ \frac{p^2}{2} + 2e^2p^2 \int d^2\xi \,
\frac{e^{ip\cdot \xi}}{\xi \cdot \xi +a^2}
 + 4e^2 \int d^2\xi \, \frac{e^{ip\cdot \xi}-1}{(\xi \cdot
\xi +a^2)^2} \right].
\label{5511}
\eeq
Notice that the notation has been simplified somehow, because in
(\ref{5511}) it must be understood that the double integration
over $d^2\xi$ is in fact a single integration on the first
coordinate $\xi_1$ and an infinite sum (one of the spatial
coordinates corresponds to the torus), exactly as in the case of
the first integration (over $p$ and $n$). After some work, one obtains
\cite{eo1}
\bea
V(R)&=& kR - \frac{(D-2)\pi}{24 R} + \frac{D-2}{2} \int dp
\sum_{n=1}^\infty \ln \left\{  4\pi e^2 \sum_{m=1}^\infty
\frac{e^{inm/R} e^{-p \sqrt{m^2+a^2}}}{\sqrt{m^2+ a^2}} \right. \label{5515}
\\ && + \left. \frac{4\pi e^2}{p^2 +n^2/R^2} \sum_{m=1}^\infty
\left[ \frac{e^{inm/R} e^{-p \sqrt{m^2+a^2}}\left( 1 +p
\sqrt{m^2+a^2}\right)-1 }{(m^2+a^2)^{3/2}} - \frac{1}{2(m^2
+a^2)^{3/2}} \right] \right\}.  \nn
\eea
No approximation is involved in Eq. (\ref{5515}). However, this
expression is quite complicated and to proceed further one has to do
some approximation, valid in the limit when $a$ is big ($a> R$). The
first approximation in $1/a$, obtained by keeping just the leading terms
of the final result, yields
\beq
V(R) \simeq kR - \frac{(D-2)\pi}{24 R} \left[ 1 - \frac{ 6\pi e^2
R^2}{a^3} \left( \gamma + 2 \ \mbox{Sinint}\, (1) - \pi  + e^{-a/(
R)} \right) - \frac{24 \pi e^2}{a (e^{a/( R)} -1 )} \right],
\label{5516}
\eeq
where $\gamma$ is Euler's constant and Sinint the standard
sinus integral (Sinint $(1)= 0.946083$). (Note that in this
expression we have kept
a couple of terms which are representative of the asymptotically smaller
contributions to the effective potential that can be dismissed
completely in this approximation.) The dependence on $a$ and $R$ could
have been ascertained by dimensional reasons. Numerically, the result
is:
\beq
V(R) = kR - \frac{(D-2)\pi}{24 R} \left\{ 1 - \frac{ e^2
R^2}{a^3} \left[ 25.3416 + {\cal O}\left( \frac{1}{a} \right) \right]
\right\}.
\label{5517}
\eeq

Having at hand this result for the effective potential, we can now
study in some detail the contribution of the antisymmetric fields  to
the
static potential. From (\ref{5517}) we see immediately that the static
potential is given by expression (\ref{558}) with a  renormalized
string tension, namely
\beq
V(R) =  k_R R - \frac{(D-2)\pi}{24 R},
\label{5518}
\eeq
where
\beq
 k_R \simeq k + 3.3172\, (D-2) \, \frac{e^2}{a^3}.
\label{5519}
\eeq
Hence, we observe that when the radius $R$ equals $R_c$, $R^2_c =
(D-2)\pi/(24 k_R)$,
it turns out that $V (R_c)=0$. The appearance of this critical radius,
$R_c$  indicates very probably ---as in more complicated string
models--- that
the quasi-static string picture ceases to be valid there, what has been
interpreted in Ref. \cite{c2} (using a different string
model as example) to be a signal for a phase transition. From
this point of view, it seems natural to interprete the effect of the
antisymmetric fields in the static potential as a renormalization of the
string constant $k$ (a one-loop correction to the classical potential),
what produces a change in the value of the critical
radius $R_c$, as compared with the one that it has in the case when
there is no coupling with antisymmetric fields.

\vspace{3mm}

\ni{\bf 5. \ Static potential and the Casimir effect
for the micro-organisms.}

In this section, we will consider the backreaction of the
velocity fields for the swimming of micro-organisms in a fluid,
especially in a two-dimensional fluid.
We use the canonical formalism
for convienience and evaluate the zero point energy
of the Hamiltonian.
The formalism is, of course, equivalent to
the path integral formalism in the previous
sections and
the obtained energy has a form similar to those of
the static potentials in the previous sections.

For the case of micro-organisms' swimming, the Reynolds number
is very
small and the Navier-Stokes equation becomes linearized.
Therefore, the equations of motion for the incompressible
fluid are given here by
\begin{eqnarray}
\label{incom}
\nabla\cdot{\bf v}&=&0, \\
\label{press}
\Delta{\bf v}&=&{1 \over \eta}\nabla p\ {\rm or}\
\Delta(\nabla\times{\bf v})=0\ .
\end{eqnarray}
Here ${\bf v}$, $p$ and $\eta$ are the velocity field,
the pressure and the viscosity coefficient, respectively.
In two dimensions, Eqs. (\ref{incom}) and
(\ref{press}) acquire the specific form
\begin{eqnarray}
\label{incom2}
\partial_z v_{\bar z}
+\partial_{\bar z}v_z&=&0, \\
\label{press2}
4\eta\partial_z\partial_{\bar z}v_{\bar z}&=&
\partial_{\bar z} p, \nonumber \\
4\eta\partial_z\partial_{\bar z}v_z&=&
\partial_z p.
\end{eqnarray}
It is an interesting point that
the equations of motion (\ref{incom2}) and (\ref{press2}) can
be derived from the following QED-like Lagrangean
in the Landau gauge:
\begin{equation}
\label{lag}
{\cal L}=2\eta
(\partial_z v_{\bar z}-\partial_{\bar z}v_z)^2
 -p(\partial_z v_{\bar z}
+\partial_{\bar z}v_z)\ .
\end{equation}
Furthermore, the Lagrangean (\ref{lag})
can be regarded as the entropy density of the fluid,
what will be shown next.

According to any standard textbook on the theory of fluid
dynamics\footnote{
We use here the notation of Landau and Lifschitz.},
one finds that the time-derivative of the fluid entropy $S$
is given by the expression
\begin{equation}
\label{entropy}
\dot S=\int dV\Bigl[ {\sigma'_{ik} \over 2T}\Bigl(
{\partial v_i \over \partial x_k}
+{\partial v_k \over \partial x_i}\Bigr)
 -{{\bf q}\cdot {\rm grad}\,T \over T^2}\Bigr]\ .
\end{equation}
Here $\sigma'_{ik}$ is the stress
tensor (in three dimensions):
\begin{equation}
\label{stress}
\sigma'_{ik}=\eta\Bigl(
{\partial v_i \over \partial x_k}
+{\partial v_k \over \partial x_i}
 -{2 \over 3}\delta_{ik}
{\partial v_l \over \partial x_l}\Bigr)
+\zeta\delta_{ik}
{\partial v_l \over \partial x_l}\ ,
\end{equation}
and ${\bf q}$ is the heat current density.
If we assume that the temperature $T$ can be globally
defined, ${\rm grad}\,T=0$, and that the fluid
is incompressible (\ref{incom}), Eq. (\ref{entropy}) can
be rewriten as
\begin{equation}
\label{entropy2}
\dot S=\int dV{\eta \over 2T}\Bigl[ \Bigl(
{\partial v_i \over \partial x_k}
 -{\partial v_k \over \partial x_i}\Bigr)^2
+p{\partial v_l \over \partial x_l}
+{\partial \over \partial x_i}\Bigl(
v_k{\partial v_i \over \partial x_k}\Bigr)\Bigr]\ . \label{29}
\end{equation}
Here we have imposed the condition of
incompressibility by means of the multiplier field $p$.
The resulting expression is equivalent to the action of
QED in the Landau gauge if we neglect the surface term.

Equation (\ref{entropy2}) tells us that the entropy of the
fluid in two dimensions satisfying the boundary condition
on the surface of the micro-organisms is given by
\begin{eqnarray}
\label{micro}
\dot{S}&=&{\eta \over 2T}\Bigl[ {i \over 2} \int dz d\bar z
\Bigl\{ 4(\partial_z v_{\bar z}-\partial_{\bar z}v_z)^2
 -p(\partial_z v_{\bar z}
+ \partial_{\bar z}v_z)\Bigr\}\nonumber \\
& & +\int_0^{2\pi}d\sigma\{ P(\sigma)(\dot Z(\sigma)
 -v^z(Z(\sigma), \bar Z(\sigma))) \nonumber \\
& & \hskip 1cm
+\bar P(\sigma)(\dot{\bar Z}(\sigma)
 -v^{\bar z}(Z(\sigma), \bar Z(\sigma))) \}\Bigr].
\end{eqnarray}
Here $\sigma$ parametrizes the surface of the micro-organism
and $Z(\sigma)$ expresses the shape of the surface, like a
string coordinate.
In the terminology of the previous sections
\[
X^0 = t = \xi_0, \ \ Z(t, \sigma)
= X^1(t, \sigma) + i X^2 (t, \sigma)
\ \ (\sigma = \xi_1 /R).
\]
The second term in Eq. (\ref{micro}) gives
the condition that there is no slipping between the surface
of the micro-organism and the fluid.

By integrating over the velocity fields $v_z$ and
$v_{\bar z}$, we obtain the following Coulomb-like
interaction between the $P(\sigma)$'s:
\begin{eqnarray}
\label{int}
\dot{S}&=&{\eta \over 2T}\int_0^{2\pi}d\sigma\{ P(\sigma)
\dot Z(\sigma)
+\bar P(\sigma)\dot{\bar Z}(\sigma)\} \nonumber \\
& & +\int_0^{2\pi}d\sigma\int_0^{2\pi}d\sigma'\{
P(\sigma)P(\sigma')G^{zz}(Z(\sigma),Z(\sigma'))
\nonumber \\
& & \hskip 1cm +\bar P(\sigma)\bar P(\sigma')
G^{\bar z \bar z}(Z(\sigma),Z(\sigma')) \nonumber \\
& & \hskip 1cm
+2P(\sigma)\bar P(\sigma')G^{z\bar z}(Z(\sigma),Z(\sigma'))
\}\ . \label{32}
\end{eqnarray}
Here $G_{zz}(z,z')$, $G_{\bar z \bar z}(z,z')$ and
$G_{z\bar z}(z,z')$ are defined by
\begin{eqnarray}
\label{green}
G_{zz}(z,z')&=&{1 \over 2\pi}{(z-z')^2 \over
|z-z'|^2}, \nonumber \\
G_{\bar z \bar z}(z,z')&=&{1 \over 2\pi}
{(\bar z-\bar z')^2 \over |z-z'|^2}, \nonumber \\
G_{z\bar z}(z,z')&=&G_{\bar z z}(z,z') \nonumber \\
&=&{1 \over 2\pi}{\rm ln}|z-z'|^2.   \label{33}
\end{eqnarray}
In order to continue integrating over $P(\sigma)$ further,
we can use the equation of motion, which is given
by the variation of $S$ with respect to $P(\sigma)$:
\begin{equation}
\label{eqofmo}
0=\dot{\bar Z}(\sigma)
+2\int_0^{2\pi}d\sigma'\{P(\sigma')G^{zz}(Z(\sigma),Z(\sigma'))
+\bar P(\sigma')G^{z\bar z}(Z(\sigma),Z(\sigma'))
\}\ .
\end{equation}
In the following, we consider the swimming motion of the
micro-organism with cilia and we assume that the shape of the
micro-organism is given by
\begin{equation}
\label{deform}
Z(\sigma)=R{\rm e}^{i\sigma}
+\sum_{n\neq 0,1}{\rm e}^{in\sigma}z_n\ ,\hskip 1cm
(z_n\ll R).
\end{equation}
We now solve Eq. (\ref{eqofmo}) with respect to
$P(\sigma)$ to  first order in $z_n$.
Then we can approximate $G_{zz}(z,z')$ and
$G_{z\bar z}(z,z')$ by
\begin{eqnarray}
\label{green2}
G_{zz}(z,z')&\sim&
 -{1 \over 2\pi}{\rm e}^{i(\sigma+\sigma')}\ ,
\nonumber \\
G_{zz}(z,z')&\sim& {1 \over 2\pi}{\rm \ln }R|1
 -{\rm e}^{i(\sigma-\sigma')}| \nonumber \\
&=&{1 \over 2\pi}\Bigl\{ {\rm \ln }R+\sum_{n=1}^\infty
{1 \over n}
({\rm e}^{in(\sigma-\sigma')}+{\rm e}^{-in(\sigma-\sigma')})\Bigr\}.
\end{eqnarray}
By expanding $P(\sigma)$ : $P(\sigma)=\sum_n{\rm e}^{in\sigma}p_n$
and substituting it into Eq. (\ref{eqofmo}), we obtain
the following equation to first order in $z_n$:
\begin{equation}
\label{eqofmo2}
0=\sum_{n\neq 0,1}{\rm e}^{in\sigma}\dot z_n
+2{\rm e}^{i\sigma}p_{-1}+2\sum_{n=1}^\infty {1 \over n}
({\rm e}^{in\sigma}\bar p_n +
{\rm e}^{-in\sigma}\bar p_{-n})\ .
\end{equation}
Then we find
\begin{eqnarray}
\label{sol}
p_n&=&-{|n| \over 2}\dot{\bar z}_{-n}, \hskip 1cm (n\neq 0,-1)
\nonumber \\
 p_{-1}&=&0\ .
\end{eqnarray}
Note that $p_0$ cannot be
determined but, actually,
they do not appear in Eq. (\ref{int}).
By substituting the solutions in Eq. (\ref{sol}), we obtain
\begin{equation}
\label{action}
\dot{S}=-{\pi\eta \over 2T}\sum_{n\neq 0,1}
|n|\dot z_n\dot{\bar z}_n\ .
\end{equation}
The \lq\lq action'' $S$ in Eq. (\ref{action}) is proportional
to the power of the micro-organism.

We now add the potential term which makes the shape of
micro-organism  to be a circle of radius $R$:
\begin{equation}
V=\lambda^2R\int_0^{2\pi}d\sigma |Z(\sigma)
 -R{\rm e}^{i\sigma}|^2=2\pi\lambda^2R\sum_{n\neq 0,1}
\bar z_n z_n \ .
\end{equation}
Then the total ``Lagrangean'' $L$ is given by
\begin{equation}
\label{lag2}
L={\pi\eta \over 2T}\sum_{n\neq 0,1}
|n|\dot z_n\dot{\bar z}_n-{2\pi\lambda^2R \over T}
\sum_{n\neq 0,1}
\bar z_n z_n\ .
\end{equation}
If we set $z_n=x_n+iy_n$, the conjugate momenta
corresponding to $x_n$ and $y_n$ are given by
\begin{equation}
\label{mom}
p_n^x\equiv {\partial L \over \partial \dot x_n}
={\pi\eta \over T}
|n|\dot x_n\ , \hskip 1cm
p_n^y\equiv {\partial L \over \partial \dot y_n}
={\pi\eta \over T} |n|\dot y_n,
\end{equation}
and the corresponding ``Hamiltonian'' is
\begin{eqnarray}
\label{hamiltonian}
H&=&{1 \over 2\pi}\sum_{n \neq 0,1}\Bigl[
{T \over |n|\eta }((p_n^x)^2+(p_n^y)^2)
+{(2\pi)^2 R \lambda^2 \over T}(x_n^2+y_n^2)\Bigr]
\\
\label{hamiltonian2}
&=&\Bigl( {4 RT \lambda^2 \over \eta} \Bigr)^{{1 \over 2}}
\sum_{n \neq 0,1}
{1 \over |n|^{{1 \over 2}}}[(a_n^x)^\dagger a_n^x
+(a_n^y)^\dagger a_n^y+1].
\end{eqnarray}
Here, the creation and annhilation operators $(a_n^x)^\dagger$,
$(a_n^y)^\dagger$, $a_n^x$ and $a_n^y$, are defined by
\begin{eqnarray}
\label{ancr}
(a_n^x)^\dagger &=&\sqrt{{T \over 4\pi R^{{1 \over 2}}\lambda
\eta^{{1 \over 2}} |n|^{{1 \over 2}}}}\Bigl\{p_n^x
+i\Bigl({4\pi^2 R \lambda^2 \eta |n| \over T}
\Bigr)^{{1 \over 2}}x_n \Bigr\},
\nonumber \\
a_n^x&=&\sqrt{{T \over 4\pi R^{{1 \over 2}}\lambda
\eta^{{1 \over 2}} |n|^{{1 \over 2}}}}\Bigl\{p_n^x
 -i\Bigl({4\pi^2 R \lambda^2 \eta |n| \over T}
\Bigr)^{{1 \over 2}}x_n \Bigr\},
\nonumber \\
(a_n^y)^\dagger &=&\sqrt{{T \over 4\pi R^{{1 \over 2}}\lambda
\eta^{{1 \over 2}} |n|^{{1 \over 2}}}}\Bigl\{p_n^y
+i\Bigl({4\pi^2 R \lambda^2 \eta |n| \over T}
\Bigr)^{{1 \over 2}}y_n \Bigr\},
\nonumber \\
a_n^y&=&\sqrt{{T \over 4\pi R^{{1 \over 2}}\lambda
\eta^{{1 \over 2}} |n|^{{1 \over 2}}}}\Bigl\{p_n^y
 -i\Bigl({4\pi^2 R \lambda^2 \eta |n| \over T}
\Bigr)^{{1 \over 2}}y_n \Bigr\}.
\end{eqnarray}
We conclude that the zero-point energy can be evaluated by means
of the formula
\begin{equation}
\label{zero}
E_0=\Bigl( {4 RT \lambda^2 \over \eta}\Bigr)^{{1 \over 2}}\sum_{n \neq 0,1}
{1 \over |n|^{{1 \over 2}}}
=\Bigl({4 RT \lambda^2\over \eta }\Bigr)^{{1 \over 2}}
\Bigl[ 2\, \zeta(-1 / 2) -1\Bigr]. \label{infra}
\end{equation}
In the second line of Eq. (\ref{zero}) we have used
the definition of the zeta function $\zeta(x)$.
Note that the apparent divergence in the zero-point
energy comes from the infrared sector. In fact,
the contribution to the zero-point energy
of the mode $n$ decreases, as $|n|$ increases. This corresponds to
the fact that the efficiency of the translation or
 rotation of the micro-organism becomes better
for higher  modes \cite{c7}.
And these infinitely decreasing zero-point energies pile
up to an infrared divergence but no contribution from an ultraviolet
 one appears.

The system of micro-organism has a symmetry of
the reparametrization of $\sigma$ in Eq.(\ref{micro})
although there is not a reparametrization symmetry
of the time $t$.
It should be noted that the zeta-function regularization
is the
regularization which keeps {\it any} symmetry
by construction since the Hamiltonian
itself is used as a regulator in the Hamiltonian
formulation.
In our case, the regularization corresponds
to the replacement
\begin{eqnarray}
\label{zetareg}
Z(\sigma)&\rightarrow&Z^s(\sigma) \nn
&\equiv & {2 \over \pi\Gamma(s+{1 \over 2})}
\int_0^\infty dt \int_{-\infty}^\infty du \int_{-\infty}^\infty dv\,
t^{2s} \e^{-v^2 + 2iut} \nn
&& \hskip 1cm \times \e^{2iuv\left({4RT\lambda^2 \over \eta}
\right)^{-{1 \over 2}}H} Z(\sigma)
\e^{-2iuv\left({4RT\lambda^2 \over \eta}\right)^{-{1 \over 2}}H} \\
\bar Z(\sigma)&\rightarrow& \bar Z^s(\sigma) \nn
(a_n^{x,y},\ (a_n^{x,y})^\dagger&\rightarrow&
|n|^{-s}a_n^{x,y},\ |n|^{-s}(a_n^{x,y})^\dagger)
\end{eqnarray}
Therefore the terms dropped by the analytic
continuation of $s$ do not break any symmetry.

 \vspace{3mm}

\ni{\bf 6. \ Summary.}

The swimming of micro-organisms in a fluid and the
motion of strings (and membranes) in  background fields
have many points in common.
If the surface of a micro-organism is viewed as a string
(or membrane), its vibration (the swimming motion)
induces a flow of the fluid.
The thus induced flow can affect the motion of
another micro-organism (another string or membrane), what results in
an interaction between the two micro-organisms,
or the motion of the original micro-organism itself, originating
a self-interaction.
Therefore, the mutual and self-interactions
between micro-organisms (represented by strings or
membranes) are mediated by the velocity field
of the fluid.

It is well-known that the interactions between strings
or membranes can be  mediated by  various background
fields, such as the graviton,  anti-symmetric fields and
 gauge fields.
In these theories, the study of how to estimate the
static potential (the Casimir energy) of strings
and membranes has been developed extensively,
incorporating  quantum effects by using, e.g., the zeta
function method.
Here, our aim has been  to apply  recently
developed techniques in  string and membrane theories
to the basic problem of micro-organisms' swimming.

In  sections 2 and 3 we have carried out the analysis of
the static potential for strings and membranes under a
constant magnetic field,
in the example of the background field $U(1)$, a gauge field,
where the existence of the magnetic field changes the
strength of the  Coulomb potential.
We have summarized the effect of the anti-symmetric field
to the static potential of the string, where by
integrating out  the anti-symmetric fields one introduces
the renormalization effect on the string tension. The case of the
coupling with antisymmetric fields has been summarized in Sect. 4.

Using this sound theoretical background, we have studied the
swimming motion of  ciliates in a two dimensional fluid.
The statistical weight of the fluid dynamics is given
by $e^S$ with its entropy $S$ (in  units of the
Bolzmann constant $k_B = 1$),
so that the \lq\lq Lagrangean" $L$ of the system should
be $-\dot{S}$.
It is  interesting to notice that the
\lq\lq Lagrangean" $L$ of our incompressible fluid is
identical to the action of QED in the  Landau gauge,
if the velocity field $v_{\mu}(x)$ is recognized as a
$U(1)$ gauge field (see Eq. (\ref{29})).
Introduction of the micro-organism, with its
surface parametrized by $X^{\mu}(\xi)$, requires the
no-slipping condition:
$\dot{X}^{\mu}(\xi)= v^{\mu}(x)|_{x=X(\xi)}$.
Then, the gauge field $v_{\mu}(x)$ can mediate the
Coulomb interaction
between the non-local momenta $P_{\mu}(\xi)$, the
Fourier transform of
$\dot{X}^{\mu}(\xi)$ (see Eqs. (\ref{32}) and (\ref{33})).

This situation of mediating a Coulomb-like interaction
between non-local  objects can be commonly found in
string (and membrane) theories coupled with
background fields.
Coupling of the gauge fields $A_{\mu}(x)$ induces a
Coulomb interaction
between the non-local currents $\partial_i X^{\mu}(\xi) $
($i=1$ in Eq. (\ref{n1}) and $i=1 ,\  2$ in Eq. (\ref{nb4})),
whereas the antisymmetric field $A_{\mu \nu}(x)$ induces
the same  interaction between the non-local surface elements
$\sigma^{\mu \nu}(\xi)= \epsilon^{ij} \partial_i X^{\mu}(\xi)
\partial_j X^{\nu}(\xi)$ (see (\ref{555})).

When we prepare two different micro-organisms, strings
or membranes,
a Coulomb-like interaction between them is generated.
This interaction is very important for
the description of phenomena involving the collective motion
of the micro-organisms,  as is the case of the red tide \cite{c12}.
Specifically, in this paper we have discussed in detail
the self-interactions by taking up a single micro-organism,
string or membrane,
the non-local configuration of which brings about
an interesting behavior on the static potential and
on the Casimir energy.
The zero-point energy of the micro-organism is found to
be smaller for the  higher vibration modes in the
ciliate swimming motion, what corresponds
to the fact that the efficiency or power becomes better
for the higher modes.

In the calculation of the Casimir energy of the micro-organism
we have found  infrared divergences rather than
 ultraviolet ones (see Eq. (\ref{infra})).
The reason for this interchange that occurs between the
ultraviolet and the infrared regions may be understood
from the fact that in the
micro-organisms' swimming, the Coulomb interaction works
between the momentum $P^{\mu}(\xi)$'s and {\it not} between
the $\dot{X}^{\mu}(\xi)$'s.
The former and the latter are Fourier conjugate with each
other, so that the ultraviolet behavior of $P^{\mu}(\xi)$
corresponds to the infrared behavior of $\dot{X}^{\mu}(\xi)$.
For such a discussion, the swimming problem of micro-organisms
in a two  (resp. three) dimensional fluid can be
dually related with the string (resp. membrane) theory
coupled with the $U(1)$ gauge field by means of the
current $\dot{X}^{\mu}(\xi)$.

\vspace{3mm}

\noindent{\bf Acknowledgments.}
We thank O. Alvarez for discussions on our approach.
This work has been supported by DGICYT (Spain), projects
PB93-0035 and SAB93-0024, by CIRIT (Generalitat de
Catalunya), and by ISF, project RI1000.

\end{document}